\documentclass[conference]{IEEEtran}
\IEEEoverridecommandlockouts
\usepackage{amsmath,amssymb,amsfonts}
\usepackage{algorithmic}
\usepackage{graphicx}
\usepackage{hyperref}
\usepackage{textcomp}
\usepackage{dirtytalk}
\usepackage{xcolor}
\usepackage{biblatex}

\begin{document}
\title{Zero-Knowledge Proofs for Questionnaire Result Verification in Smart Contracts}

\author{\IEEEauthorblockN{1\textsuperscript{st} Carlos Efrain Quintero-Narvaez}
\IEEEauthorblockA{\textit{School of Science and Engineering} \\
\textit{Tecnologico de Monterrey}\\
Monterrey, Mexico \\
efrainq@tec.mx}
\and
\IEEEauthorblockN{2\textsuperscript{nd} Raúl Monroy-Borja}
\IEEEauthorblockA{\textit{School of Science and Engineering} \\
\textit{Tecnologico de Monterrey}\\
Monterrey, Mexico \\
raulm@tec.mx}
}

\maketitle

\begin{abstract}
We present an implementation of a Web3 platform that leverages the Groth16 Zero-Knowledge Proof schema to verify the validity of questionnaire results within Smart Contracts. Our approach ensures that the answer key of the questionnaire remains undisclosed throughout the verification process, while ensuring that the evaluation is done fairly. To accomplish this, users respond to a series of questions, and their answers are encoded and securely transmitted to a hidden backend. The backend then performs an evaluation of the user's answers, generating the overall result of the questionnaire. Additionally, it generates a Zero-Knowledge Proof, attesting that the answers were appropriately evaluated against a valid set of constraints. Next, the user submits their result along with the proof to a Smart Contract, which verifies their validity and issues a non-fungible token (NFT) as an attestation of the user's test result. In this research, we implemented the Zero-Knowledge functionality using Circom 2 and deployed the Smart Contract using Solidity, thereby showcasing a practical and secure solution for questionnaire validity verification in the context of Smart Contracts.
\end{abstract}

\begin{IEEEkeywords}
zero-knowledge, proof, web3, circom, questionnaire, solidity, smart contract, blockchain
\end{IEEEkeywords}

\section{Introduction}

 Blockchain technologies and Smart Contracts have gained significant traction recently, offering transparent and decentralized solutions to problems in different domains. Doing trustless verification and attestation of facts while preserving privacy is one of these domains. In particular, verification of questionnaire evaluation results presents a unique challenge, as this calls for preserving confidentiality of privileged information such as the answers to the questionnaire. This paper introduces an implementation of a Web3 platform that leverages Zero-Knowledge Proofs (ZKPs) and Smart Contracts to address this challenge.

 By employing the Groth16 Zero-Knowledge Proof schema, we establish a robust framework that enables users to receive proofs of the validity of their test results without revealing the underlying evaluation methods or answers. Furthermore, users can then produce attestations of their results in the form of Non-Fungible Tokens (NFTs) generated by a Smart Contract that verifies the ZKP we just mentioned. This way, we obtain an end-to-end questionnaire answering, validating and attesting protocol.

The significance of this research lies in enhancing trust and reliability in the evaluation of questionnaire-based assessments. By leveraging ZKPs in this way, test appliers can securely show that their evaluations are fair without compromising the integrity of the test. In turn, users can obtain reliable evidence of their performance in these assessments and share it with third parties without worrying about suspicions of forgery or other dishonest practices.

We provide a comprehensive overview of our implementation, including the design and architecture of the Smart Contracts using Solidity and the implementation details using Circom 2 for the ZKP functionality. Additionally, we present the evaluation results, discussing the system's performance, security, and privacy aspects.

By introducing this innovative approach to questionnaire validity verification in smart-contracts, we aim to open avenues for integrity-preserving attestation mechanisms in a range of applications. The subsequent sections delve into the background, methods, results, and discussion, highlighting the significance and potential implications of our research, as well as future directions for improvement.

\section{Background}
\subsection{Blockchain and Smart Contracts}
Blockchain technologies are being increasingly adopted in different domains for their features regarding transparency and decentralization. Bitcoin \cite{nakamoto2009bitcoin} was the first proposed protocol based on this architecture, combining in its functioning techniques such as Proof-of-Work (PoW) \cite{back2002hashcash}, SHA-256 hashing, Merkle Trees, and the Elliptic Curve Digital Signature Algorithm (ECDSA). It is specially powerful because of its decentralized nature, provided through the Proof-of-Work validation done with each block of transactions added to a public registry known as the Blockchain. In the following years, variations on the protocol proposed by Bitcoin appeared, collectively known as cryptocurrencies, each having their own blockchain where the transactions are registered. Ethereum \cite{Buterin2013}, Monero \cite{sabarhagen2013}, Polkadot \cite{wood2020}, Polygon \cite{kanani2020}, NEAR, Solana \cite{solana2020} and Avalanche \cite{Rocket2019ScalableAP} are some examples of these derivate protocols.

Of the protocols we just mentioned, Ethereum is a critical one to talk about as it was the first to work with the Ethereum Virtual Machine (EVM) \cite{Buterin2013} architecture, an essential feature on which many other blockchains are built upon. The EVM allows protocols to execute code in a decentralized manner and register all of its execution steps on the blockchain, making the results of that code transparent and unbiased. This code is then deployed to an \textit{address} stored in the blockchain and referred to as a Smart Contract, i.e. a contract that enforces itself automatically. 

\begin{figure}
    \centering
    \includegraphics[width=0.45\textwidth]{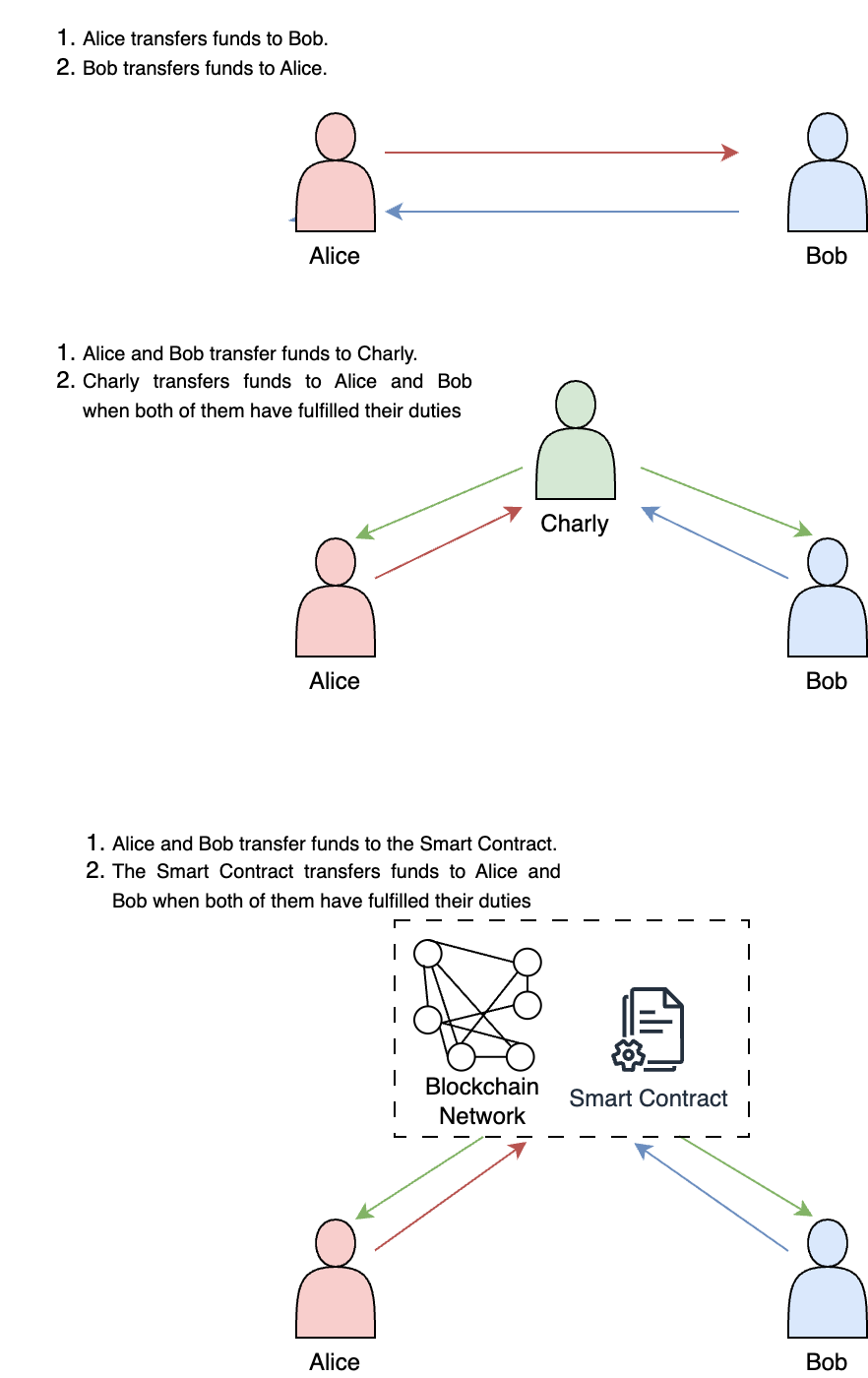}
    \caption{Diagram of three different protocols for exchanging two different currencies. From top to bottom, in the first one Alice transfers to Bob first, meaning she has to trust Bob not to keep her funds and without fulfilling his part of the deal. In the second one, Charly, a third party is involved having the task of enforcing that both Alice and Bob fulfill their parts of the deal, meaning now both Alice and Bob have to trust Charly to enforce correctly. In the third diagram, Charly is replaced by a Smart Contract executed on a blockchain network, meaning now Alice and Bob do not have to place trust on anyone except on the stability of the network.}
    \label{fig:smartcontractdiagram}
\end{figure}

A simple example of a task that can be performed using a Smart Contract is that of currency exchange. Usually, when two entities, Alice and Bob, each holding a different digital currency, want to exchange one for the other at a certain rate, a level of trust is needed. Indeed, either Alice has to transfer its currency to Bob first or vice versa, making the first entity that transfers vulnerable to the other not fulfilling their part of the deal. One way to balance this trust is by introducing a third-party entity, Charly, to which both parties transfer their funds so that Charly is tasked with ensuring that the deal is fulfilled. However, this only changes the protocol so that now both Alice and Bob have to trust Charly to act correctly. Here is where Smart Contracts come in, instead of having the parties transfer to a third one, transfer to a Smart Contract entity that automatically verifies that the deal is fulfilled without Alice and Bob having to trust in it, as it is executed on a decentralized network. This is illustrated in \autoref{fig:smartcontractdiagram}. These trustless features of Smart Contracts then lead the way to a variety of applications to be built.

Decentralized Finance (DeFi) \cite{Aave}, Automated Market Makers \cite{uniswap}, and decentralized social media \cite{Lens-Protocol}, are among the many uses of Smart Contracts that have developed in recent years. Once again, the decentralized and trustless nature in the execution of the code for each of these applications is what allows them to provide an enhanced trustless performance in comparison to their traditional counterparts.

As we have said before, the essential base on which the Smart Contract infrastructure is built is the Blockchain. Let us note some of the implications of one particular attribute of the Blockchain, its functioning as a public registry. Every time a new block of valid transactions appears, it needs to be \say{mined}, i.e. finding a valid \say{nonce} value that when added at the end of the transaction block makes its SHA-256 hash have certain characteristics. These characteristics are often that the hash has a number of zeros at the beginning, a number determined by the \say{difficulty} imposed by the network. This mining process is non-trivial by design, so that \say{work} needs to be performed before adding any transaction to the public ledger, hence the Proof-of-Work name. Notice that in order for the network to reach consensus that the work has been performed and the transactions validated accordingly, it is necessary for everyone to have complete access to each new transaction block. This makes any transfer of funds in the Blockchain public, including its receivers and senders. For this reason, it would be convenient to have a method through which the network could validate transactions without receiving compromising knowledge about its users, this is where Zero-Knowledge Proofs come in to play.

\subsection{Zero-Knowledge Proofs}
In the context of Blockchain transactions, one problem that arises is that, due to the public consensus nature of the protocol, all transactions are public. It is straightforward to check the movements of an entities funds by just checking the Blockchain registry. Although this can be partially avoided due to the easiness with which one can create new addresses (if an entity’s address is disclosed, it can just send its funds to a new one), it is still possible to just trace back the origins of any funds. A possible solution to this could work if there was a way for the network to verify the validity of transactions without them knowing the sender or the receiver, i.e. if there was a way to prove that a transaction is valid without disclosing any additional knowledge. Indeed, Zero-Knowledge Proofs are capable of doing this, as we will explaint next.

A Zero-Knowledge Proof (ZKP) is a method through which one entity can prove a statement to another one without disclosing more information than the fact that particular statement is true. For example, given a fixed function $f:X\longrightarrow Y$ and a fixed value $y\in Y$. If we wanted to prove to a third party that we know a value $x\in X$ such that $f(x)=y$, a trivial solution would be to just disclose the value of $x$ we know. However, that shares more information than the fact that the statement is true. A ZKP would entail generating a proof generator function $P$ and a proof verifier $V$. Such that we can generate a \textit{witness} $w=P(x)$ that does not disclose the value $x$, and that another entity can then evaluate with the verifier function $V(w)$ and obtain a true or false value that determines whether the \textit{witness} was created from a valid solution or not. This way, we can prove to another entity that we know a value of $x$ such that $f(x)=y$, while disclosing zero knowledge about which specific value of $x$ we know.

For a more intuitive example, consider a hashing function $\mathrm{hash}: I \longrightarrow H$ from an input space $I$ to a hash space $H$, this function could be SHA-256, SHA-1, etc. Also, consider two entities, Alice $A$ and Bob $B$. Imagine that Alice guards an entrance to a treasure cave, to which one can only enter by having a password $\rho$ that is verified using its hash, i.e. Alice has the hash of the password $h = \mathrm{hash}(\rho)$. However, Alice herself does not have the password and cannot enter the cave. Now consider that Bob does have the password $\rho$ and wants to enter the cave, but he does not want to disclose the password to Alice as to not have to share it with her. For most use cases, this can be achieved by having Bob compute the hash of $\rho$, and show it to Alice, having her verify that it is the same as $h$, the one she has. However, say that $h$ is not disclosed to Alice through a private channel but through some public method as to ensure that she verifies passwords correctly and does not commit any dishonesty. This renders just showing the hash value $h$ of the password not a valid method for verification. Usually, the only method left would be for Bob to disclose $\rho$ to Alice and have her compute its hash, but there is another way to do this, which involves Zero-Knowledge Proofs.

Groth16\cite{groth16} is a Zero-Knowledge Proof system designed to generate a proof-verifier pairing schema for satisfiability of any \textit{arithmetic circuit}. What is important about the arithmetic circuit satisfiability problem is the fact that it is NP-Complete and thus allows for a wide arrange of statements to be translated into that form, including those which are often of interest in the context of ZKPs. Note that, arithmetic circuits are comprised of gates that compute arithmetic operations (addition and multiplication) on a field $\mathbb{F}$, with wires connecting the gates to represent results from one going to another. Thus, recall the SAT problem, which is NP-Complete and formulated as 
\say{given a statement with boolean variables, is it possible to find an assignment for them such that the statement is true?}. Similarly, the \textit{arithmetic circuit satisfiability} problem is NP-Complete too and formulated as \say{given a statement with variables with values from a field $\mathbb{F}$, is it possible to find an assignment for them such that the statement is true?}. This makes Groth16 an ideal system for working when generating ZKPs, being used for ZK languages like Circom 2 \cite{Muñoz-Tapia2022}.

Coming back to our conundrum with Alice and Bob in the treasure cave. As it happens, hashing functions such as SHA-256 can be expressed in terms of arithmetic circuit satisfiability and can thus have a corresponding Zero-Knowledge Proof generated. Therefore, in the situation we had, it is enough for Alice to have a SHA-256 ZK proof verifier $V$ and for Bob to use the prover function $P$ to generate a witness $w=P(\rho)$. Bob then shares $w$ with Alice, so she evaluates it with the verifier $V(w)$ and determines that the witness is valid. Thus, Alice can be sure that Bob does have the password without receiving any information about what is its exact value.

Finally, it is straightforward to see how this can be applied to Blockchain transaction verification. Suppose that we have a transaction $t$ represented in some space $T$, along with a validation function $v$ that tells whether $t$ is valid with the current state of the Blockchain, having $0$ or $1$ as output. Then for the transaction to be added by a miner into the registry, it would suffice to send a proof that $v(t) = 1$. Then, we could translate the computation of $v$ into the Groth16 schema by using a compiler such as Circom 2.This would then allow us to generate a proof-verifier pair $P$ and $V$. And finally, as was done with the Alice and Bob case, the sender would just generate a witness of the transaction $w=P(t)$ and send it to the miners which would verify it by evaluating $V(w)$. This schema is used by cryptocurrencies such as Monero and Zcash. 

Now, see that in the context that concerns this paper, a ZKP could be generated for the validity of the evaluation of a questionnaire according to some set of constraints. This would allow the user or any third party to get a convincing proof of the evaluation having been performed fairly, without revealing any information about the answers that give certain results.

\section{Design}
Our implementation makes heavy usage of Circom 2 and its features for generating Solidity code for ZK proof verifiers. This code is then integrated into an ERC-721 Smart Contract that allows the user to mint an NFT when a result of the questionnaire with the corresponding proof is provided. A basic implementation was made for a  new Web3 platform called P3rsonalities, intended to have a personality test with results validated with a ZK proof and attested through a generated ERC-721 NFT. The deployed contract code with the Circom 2 generated Solidity verifier can be found at the \href{https://anonymous.4open.science/r/P3rsonalities}{P3rsonalities GitHub Repository}.

We designed the Circom 2 code in such a way that it receives a two bit masks and one integer as inputs, one bit mask representing the user's answers for each question, the other representing the answer key,  and the integer representing the result of the test, having a total of $10$ questions. The questions are divided into two groups so that each one represents an attribute of the final result, i.e. the final result will be a two bits integer, each bit representing an attribute. After compiling, Circom 2 generates two files we use, one WebAssembly script for the generation of the ZK witness and another Solidity script for verification of said witness on a Smart Contract executed on an EVM Blockchain. It is important to note that the generated Solidity code makes heavy use of the assembly functionalities available on the EVM, as to ensure that gas costs for executing the verification are as low as possible.

The witness generator along with the questionnaire evaluator are deployed to a centralized server, in this case an AWS Lambda function, for easy deployment and access from a REST API endpoint. The user then makes a request to this API, sending its answers to the questionnaire. The API then executes the witness generator code and returns the result of the test together with the ZK witness.

The user then makes a call, with the generated witness as an input, to the deployed ERC-721 Smart Contract for minting the NFT, attesting the result returned by the API. This Smart Contract is modified so that the Solidity verifier script generated by Circom 2 is used as a required check before minting the NFT to the user's address. 

At the end of the procedure, the user obtains a Soulbound NFT (an NFT that cannot be transferred) representing the result of their answers to the questionnaire. As this NFT can only be generated when the user possesses a valid witness for that result, it holds a special value as evidence for anyone interested in verifying the results of such a test.
\begin{figure}[htbp]
    \centering
    \includegraphics[width=0.45\textwidth]{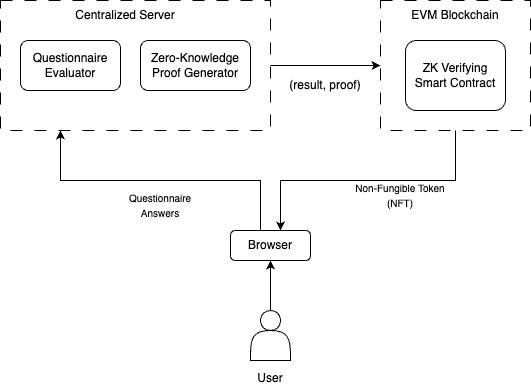}
    \caption{Architecture diagram of the complete Web3 platform for questionnaire verification generation.}
    \label{fig:enter-label}
\end{figure}

\section{Discussion}
The significance of the use of the described approach for the P3rsonalities platform lies in the resulting Soulbound NFT. This NFT is unique and cannot be transferred, making it a valuable piece of evidence for anyone interested in verifying the results of the personality test. By combining ZK proofs, ERC-721 NFTs, and the Circom 2 generated Solidity verifier, we have created a secure and tamper-evident solution for result validation and attestation.

Our implementation demonstrates the practical application of ZK proofs and NFTs in the context of questionnaire evaluation. The use of Circom 2 and its integration with ERC-721 Smart Contracts allowed us to achieve efficient and secure result validation. The results provide an immutable and verifiable record of the user's test result, which holds value for various purposes, such as research and identity verification.

Further research and experimentation can explore scalability, performance optimization, and potential extensions of this approach. Furthermore, it is also worth it to research the potential vulnerabilities of the protocol implemented here. Indeed, although this protocol is good at first glance, it still has some inherent vulnerabilities, such as users being able to \say{cheat} by sharing the answers that led to a certain result of the test in the past. However, this issue can be resolved by having a larger bank of questions such that it is improbable for two users to get the same set of questions. Thus making it infeasible to cheat by sharing the answers to an instance of the test.

On the matter of scalability, this protocol is mainly limited by the capacity of the centralized server that executes the evaluation of the questionnaire and generates the corresponding proof. On the contrary, the verification part of the protocol executed on the EVM blockchain scales pretty well in comparison, due to the decentralized nature of the blockchain network. A possible solution to the bottleneck caused by the centralized server could be to decentralize that part too. However, this is difficult as executing the questionnaire evaluation on the blockchain requires publishing the corresponding code, compromising the integrity of the evaluation. Decentralized access control protocols like Lit Protocol \cite{lit} could offer a solution to this issue, but further research is needed.

\section{Conclusions}
In conclusion, our implementation showcases the successful utilization of ZK proofs and ERC-721 NFTs to create a robust and tamper-evident system for result validation and attestation in the context of questionnaire evaluation. By integrating Circom 2's features for generating Solidity code and integrating it into our NFT, we achieved efficient verification of user-provided answers using ZK proofs, resulting in the minting of a Soulbound NFT that serves as immutable evidence of the test results. This approach holds promise not only for personality tests, as we did here, but also for various applications requiring secure result validation and attestations. Further research can explore scalability, optimization, and potential vulnerabilities, in order to expand the usability of this solution across different Web3 platforms and domains beyond questionnaire evaluation.

\printbibliography
\end{document}